\documentclass[twoside,final]{pro12}

\usepackage{graphicx}
\usepackage[singlelinecheck=false 
]{caption}

\usepackage{epsfig}
\usepackage{epstopdf} 

\usepackage[ansinew]{inputenc}
\usepackage{amsmath}

\head{Ch. Helling and I. Vorgul} {Atmospheres of extrasolar planets} \label{Hellingpaper}

\begin{document}

\title{Insight into atmospheres of extrasolar planets through plasma processes}
\author{Ch. Helling\adress{Centre for Exoplanet Science, University of St Andrews, UK} ~and I. Vorgul$^*$}


\maketitle \thispagestyle{empty}

\begin{abstract}

Extrasolar planets appear in a chemical diversity unseen in our own solar system. Despite their atmospheres being cold, continuous and transient plasma processes do affect these atmosphere where clouds form with great efficiency. Clouds can be very dynamic due to winds for example in highly irradiated planets like HD\,189733b, and lightning may emerge. Lightning, and discharge events in general, leave spectral fingerprints, for example due to the formation of HCN. During the interaction, lightning or other flash--ionisation events also change the electromagnetic field of a coherent, high energy emission  which results a characteristic damping of the initial, unperturbed  (e.g. cyclotron emission) radiation beam. We summarise this  as 'recipe for observers'. External ionisation by X-ray or UV e.g. from within the interstellar medium or from a white dwarf companion will introduce additional ionisation leading  to the formation of a chromosphere. Signatures of plasma processes therefore allow for an alternative way to study atmospheres of extrasolar planets and brown dwarfs.
\end{abstract}



\section{Introduction}

\begin{figure}[ht]
\centering
\includegraphics[width=9cm, angle=-90]{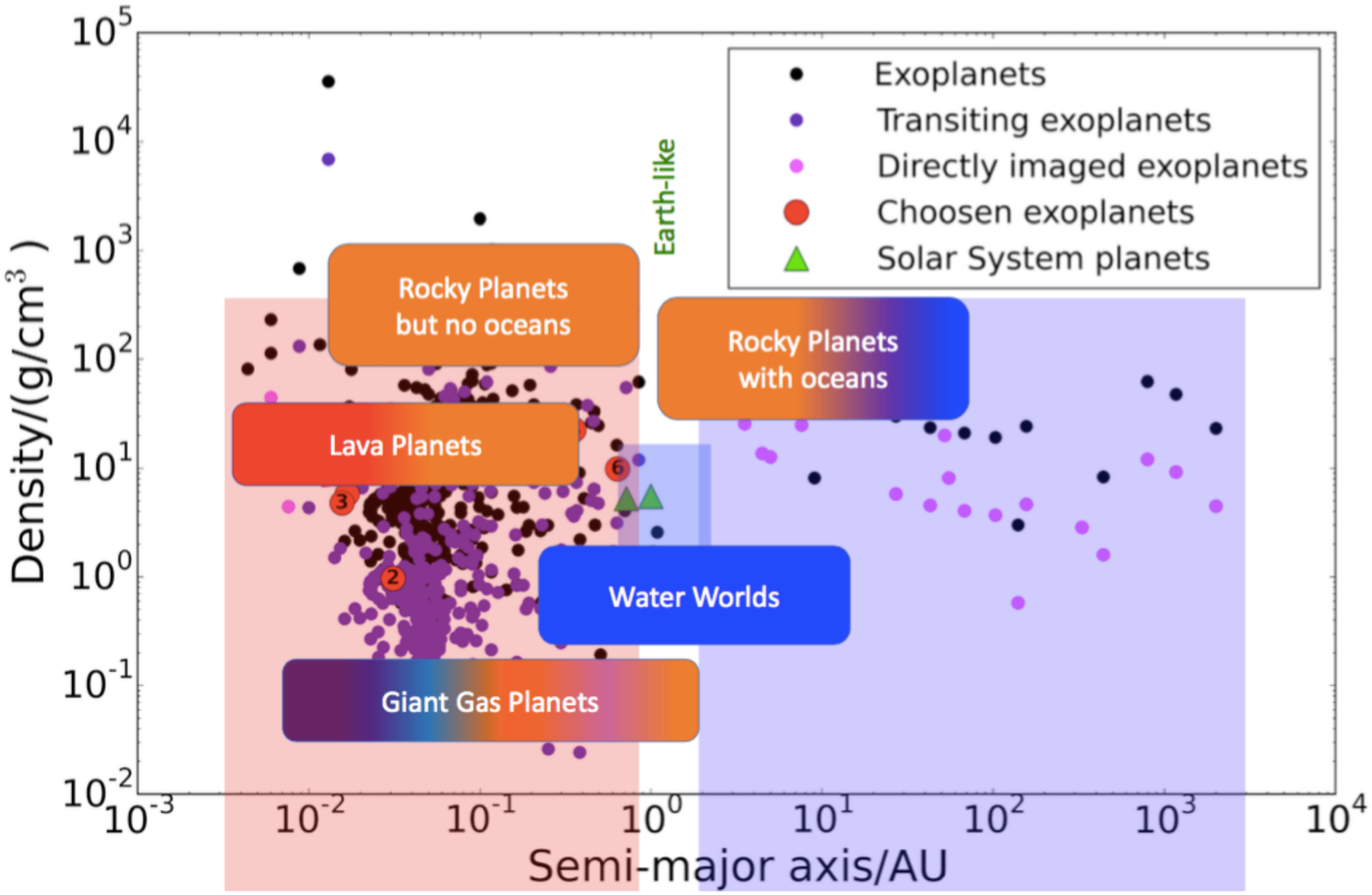}
\caption{More than 3000 extrasolar planets are known. Bulk
  densities, as in this plot, show that their diversity reaches from
  Giant Gas Planets (low bulk density) to lava planets, water worlds,
  or rocky planets with or without oceans (high bulk density). The
  transit method allows to discover planets that are very nearby their
  host star while the direct imaging method has shown that extrasolar
  planets can also exists rather far away from their host star. This
  diversity will also be reflected by the planets atmosphere.}
\label{fig:div}
\end{figure}

Extrasolar planets are discovered at an increasing rate and in an
unexpected diversity (Fig.~\ref{fig:div}). Now, observational efforts
move towards analysing these planets across the whole electromagnetic
spectrum. Optical and near-infrared observations begin to provide
insight into the chemical composition of exoplanet atmospheres. Radio observation will allow to study the plasma
processes involved. Though radio emission is typically observed from
hot stellar plasma in astrophysics, any source of electrons accelerated in a magnetic field will create radio emission. Hallinan et al. [2008] and  also  Williams et al. [2015] demonstrated that radio emission  emerges from brown dwarfs which have similarly cool
temperatures like giant gas planets (e.g. Jupiter, Saturn,
HD\,189733b). Based on these observations, brown dwarfs are suggested to have very strong
magnetic fields of the order of 1000~G. Hallinan et al. [2015] proposed
that the radio emission observed on the late M-dwarf LSR J1835
+ 3259 could be caused by a super--aurora. The dissipated energy is $10^4$ times larger than that produced in the Jovian magnetosphere, and a magnetic field strength of 1550G$\,\ldots\,$2850G is inferred.

The emergence and the efficiency of plasma processes inside
atmospheres is determined by the local degree of ionisation which
directly determines the atmospheric electric conductivity. The local degree of
ionisation is affected by the local thermal temperature, and by other
atmospheric processes like strong winds (Alfven ionisation) or
lightning.  External cosmic ray or X-ray irradiation will affect
the gas ionisation, too, and hence link local plasma processes inside
the atmosphere to environmental effects that are determined by the planet's host
star and planetary system environment [e.g. Helling et al., 2016a,b; Longstaff et al. 2017]. Transient  events (lightning), and continuous processes (external irradiation) can occur.
Plasma processes can furthermore affect the local chemistry (e.g. lightning is natural source for  NO$_{\textrm x}$ on Earth), and hence, they may appear as spectral fingerprints in the observed optical and near-IR spectra [e.g. Rimmer et al. 2014; Ardaseva et al. 2017]. A possible spectral fingerprint for lightning on extrasolar planets  maybe HCN [Hodosan et al., 2016].

Space and ground based observations [e.g. Sing et al., 2016; Nikolov et
al., 2016] demonstrate that brown dwarfs and extrasolar planets are
covered by clouds. Kinetic cloud formation models had predicted the formation of cloud particles with a wide variety of sizes and made of a mix of materials in brown dwarfs and extrasolar planets [Helling et al., 2008]. Lee et al. [2016] included the kinetic cloud modelling into  globally circulating atmospheres for
HD\,189733b and demonstrated the chemical diversity of clouds across the planetary globe. These simulations demonstrate that
the atmosphere of the highly irradiated giant gas planet HD\,189733b
is filled with mineral cloud particles that either are small enough
to follow the hydrodynamic wind motion due to frictional coupling, or
cloud particles are big enough to frictionally decouple and fall into
deeper atmospheric layers where their growth by chemical surface reactions is amplified due to
higher densities (hence increased collision rates with the gas). Both cases suggest that cloud particles are charged
due to triboelectric processes. Winds and gravitational settling provide mechanisms for large-scale charge separation,  which then can lead to the emergence of lightning in these extrasolar atmospheres [see Hodosan et al., this issue]. 

\section{Magnetic coupling in weakly ionised atmospheres}

The collisionaly dominated part of a brown dwarf/ giant gas planet
atmosphere is rather cold for yielding a substantial degree, $f_{\textrm c}$,  of thermal
ionisation. The maximum values of $f_{\textrm c}=p_{\textrm e}/p_{\textrm gas}
\approx 10^{-7}$ ($p_{\textrm e}$: electron pressure, $p_{\textrm gas}$: total gas pressure) occur at the top and bottom of the atmosphere
[Rodriguez--Barrera et al., 2015]. The main electron donors are K$^+$
and Na$^+$, followed by Mg$^+$ and Fe$^+$. 

\begin{figure}[ht]
\centering
\includegraphics[width=13cm]{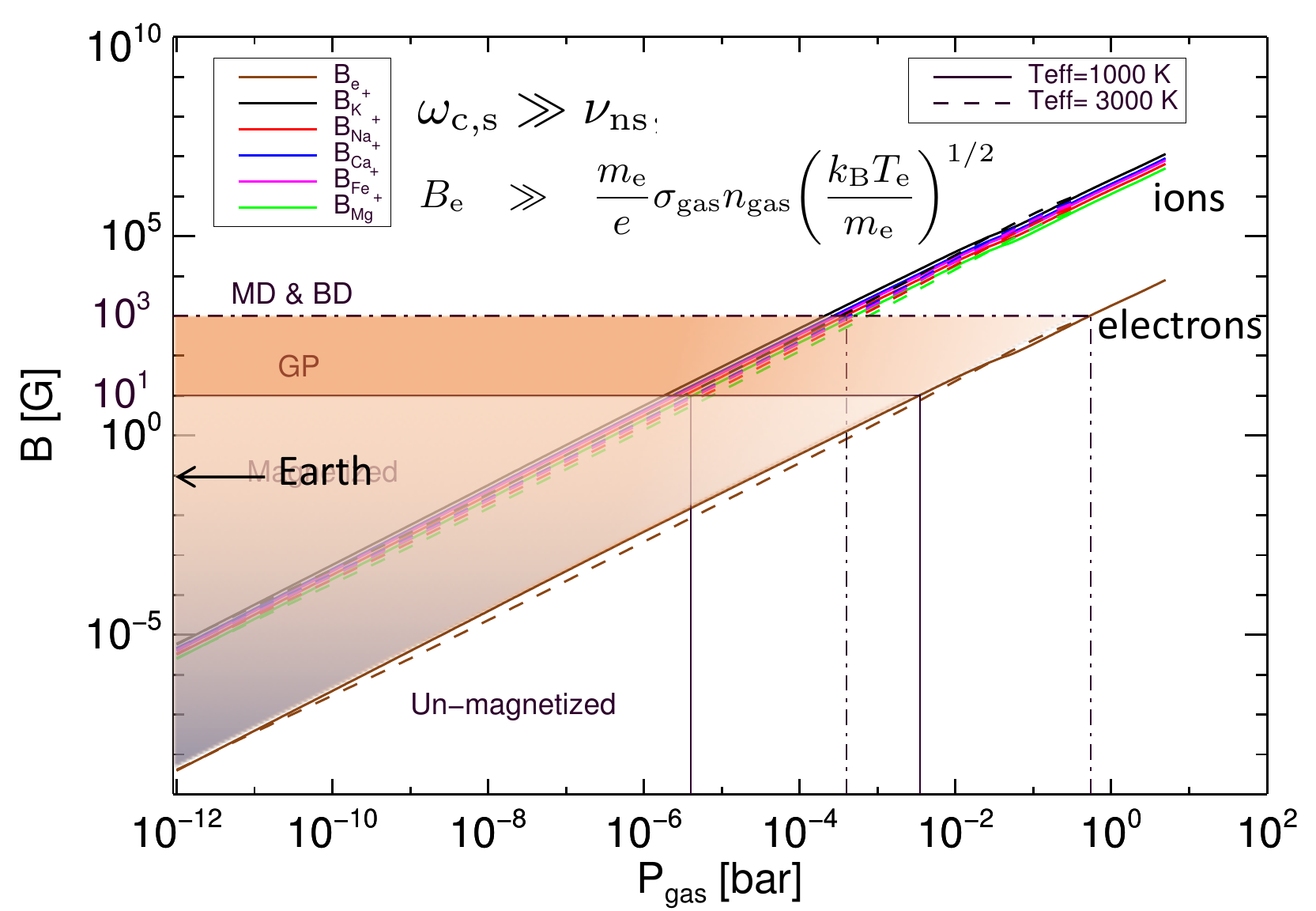}
\caption{The critical magnetic field strength, $B_{\textrm e}$ [G]
  above which charged particle's motion will be dictated by an ambient
  magnetic field. The plot demonstrates that the upper atmosphere of
  objects with T$_{\textrm eff}$=1000~K (solid lines) and T$_{\textrm
    eff}$=3000~K (dashed lines) can be magnetically coupled already in
  the case of thermal ionisation. Horizontal lines indicate boundaries
  for the shaded areas for M-dwarf/Brown Dwarfs and (MD \& BD) and
  giant gas planets (GP). The closeness of the colour lines shows that the specifics of the electron donors do not change the results significantly.}
\label{fig:Be}
\end{figure}

Rodriguez--Barrera et al. [2015] used the cyclotron and the
collisional frequencies to derive a maximum local magnetic field
strength that would be required to achieve magnetic coupling of the
local gas with $T_{\textrm e}=T_{\textrm gas}$, $$B_{\textrm e}\gg (m_{\textrm e}/c)\sigma_{\textrm
  gas}n_{\textrm gas}[(k_{\textrm B}T_{\textrm e})/(m_{\textrm
    e})]^{1/2}$$ ($T_{\textrm e}$: electron temperature, $T_{\textrm gas}$: gas temperature, $m_{\textrm e}$: electron mass, $c$: speed of light, $\sigma_{\textrm gas}$: collisional cross section, $n_{\textrm gas}$: gas number density, $k_{\textrm B}$:  Bolzmann constant). The values of
$B_{\textrm e}$ indicate where the charged particle's motion will be
dictated by an ambient magnetic field. Figure~\ref{fig:Be}
demonstrates that the upper parts of brown dwarf and giant planet
atmospheres can be magnetically coupled. The
critical magnetic field strength is higher for ions than for
electrons as shown in Fig.~\ref{fig:Be}.

Our findings suggest that non-irradiated brown dwarfs (and their giant
gas planet counterparts) can develop a chromosphere as result of
magnetic coupling of the upper atmosphere  that will allow for e.g. Alfven wave
heating. Systematic H$\alpha$ observations as presented by Schmidt et al. [2015]
support this idea. We note that a chromosphere on brown dwarfs may not necessarily form from wave heating alone, but that external irradiation in form of high-energy X-ray in the ISM, from a white--dwarf companion and  inside a star-forming region may ionise the upper atmosphere completely [Rodriguez--Barrera, 2016]. This becomes of particular interest for White--dwarf/Brown Dwarf binaries like the irradiated brown dwarf WD0137-349B [Casewell et al., 2015; Longstaff et al., 2017]. The high-energy radiation field of the white-dwarf primary does ionise the upper atmosphere of the companion brown dwarf which has been observed to  emit in He, Na, Mg, Si, K, Ca, Ti and Fe. Consequently, the authors suggest that this brown dwarf most likely exhibits a chromosphere.

\section{The fingerprint of lightning in a coherent emission beam}

\begin{figure}[ht]
\centering
\includegraphics[width=13cm]{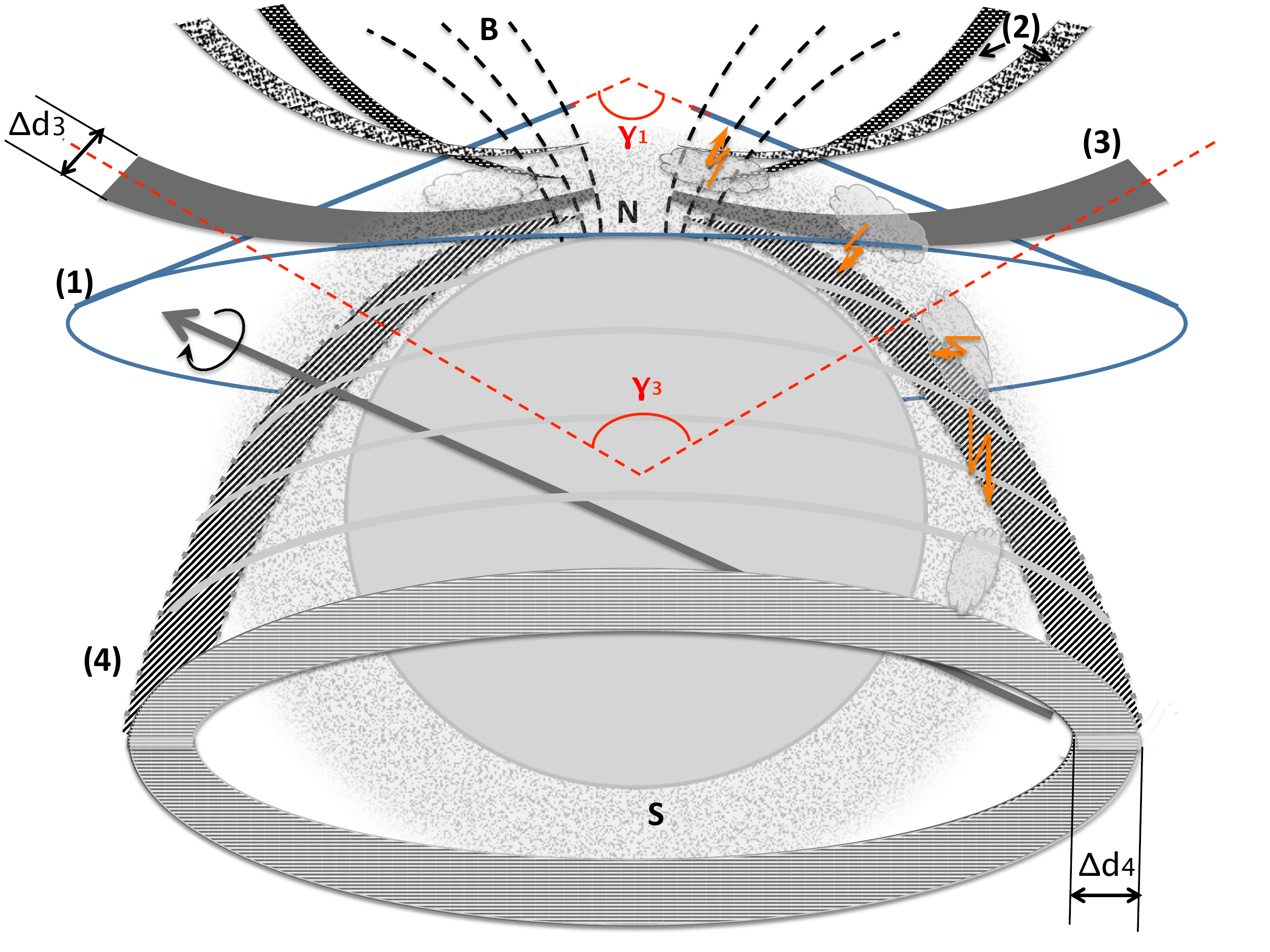}
\caption{Coherent (radio) emission interacting with a cloudy
  atmosphere where lightning occurs. Depending on the source area of
  the coherent emission near the magnetic pole, the emission cones (1
  -- 4) have different opening angles ($\gamma$). The beam will
  interact more ($\Delta d_4$) or less ($\Delta d$ for one of cones 2) with the ambient
  medium, and even beam crossing may occur (2).}
\label{fig:cones}
\end{figure}

Brown dwarfs are observed to emit coherent radio emission which was
interpreted as cyclotron maser emission [Hallinan et al., 2006].  The cyclotron radiation
emerges nearly perpendicular to the magnetic field lines, but it will be 
refracted by the atmospheric material.  The refracted radio beam may
then cross atmospheric regions where lightning occurs (orange arrows in Fig.~\ref{fig:cones}), and the lightning will effect the beam's electromagnetic field such that it leaves a
time-signature on it [Vorgul \& Helling, 2016]. The geometry of this is
shown in Fig.~\ref{fig:cones}. Examples for the different
emission cone cases are discussed in Vorgul \& Helling [2016], 
including CU Virgines showing a two--beam crossing due to the refraction
of the atmosphere (cones 2), Neptune with $\gamma=80^o$ and $\Delta
d=15^o$ (e.g. cone 3) or the Brown Dwarf NLTT 33370AB with  $\gamma=35^o$ and $\Delta
d=20^o-23^o$ (e.g. cone 4).

Cyclotron maser emission (CME) can occur at relatively low plasma density only, when the electron plasma frequency is significantly smaller than the electron cyclotron frequency. Terrestrial CME originates from elongated plasma cavities along magnetic field lines [Pritchett et al.  2002, Burinskaya \& Shevelev 2017].  The source regions known as auroral cavities are of relatively low density compared to their surroundings.  This low density allows the plasma frequency for the electrons to be much smaller than their cyclotron frequency. Figures 3 and 11 in [Rodriguez--Barrera et al. 2015] suggest that this is the case also in brown dwarf and giant gas planet atmospheres.
 The emission from such cavities escapes due to gradual inhomogeneity of the surrounding plasma, for which a connecting branch between the X-mode (at which the emission is generated) and a vacuum propagating mode appears in dispersion curves [Cairns et al. 2008]. Expanding their consideration of the local dispersion relation, Cairns et al. [2011] show how global growing eigenmodes can be constructed for that scenario.

The cyclotron emission produces cavities that extend from 1000 km to 8000 km altitude in the Earth magnetosphere, but the most powerful emission occurs  near the low-altitude limit  [Gray 2012].   We suggest to  probe a lightning-affected region with such a cyclotron emission. 
While most of the ground-to-cloud lightning discharges are initiated below the altitude at which CME originates, 
the interaction does not require the radiation crossing exactly the lightning channel, but the larger lightning-affected regions.  Mallios et al [2012] investigate charge transfer  before, during and after a cloud-to-ground lightning discharge.  The charges are transferred both from the cloud to the ground and from the cloud to the ionosphere. Mallios et al [2012] show that  the amount of charge that is transferred to the ground, due to currents flowing in the vicinity of the thundercloud during a transient time period following a lightning discharge (cloud to ground), can be several times smaller than the amount of charge that is transferred to the ionosphere during the same time period. The high amount of charges can then be transferred to an Earth altitude from 2 km to 13 km, hence producing a quick rise of conductivity far away from the initial event.  These are altitudes well matching those where the cyclotron emission occurring on Earth at the low-altitude limit can propagate through, and so this sudden change in conductivity can be probed. It is then likely by analogy, that similar scenarios occur on other planets and also on brown dwarfs. 
 Cyclotron radio signals from BDs come widened (large $\Delta d$ in Fig.~\ref{fig:cones}) compared to those from stars, which suggests that the emission is disturbed by propagation effects. This is likely to occur when CME is emitted in cavities and then escapes through an inhomogeneous plasma. Thus, we suggest to expect  that cyclotron emission can probe the high-gradient conductivity rise due to lightning discharges. We point out that this idea of lightning affecting radio waves originating from other processes or sources, is not entirely new as e.g. Haldoupis et al. [2004], Inan \& Cummer [2010] and others investigate the impact of lightning induced TLE disturbances on the conductivity profile in the Earth mesosphere
by tracking man-made radio signals.

The idea that lightning affects the signature of radio emission has
also been put forward by Schellart et al. [2015]. Schellart et
al. demonstrated that terrestrial lightning leaves a fingerprint on
the radio emission from cosmic ray induced air showers. Vorgul and
Helling [2016] developed the mathematical framework for
the effect of extraterrestrial lightning on coherent radio emission on
brown dwarfs: They developed a general model to describe the influence of flash ionization
events (like lightning or expositions of similar intensity) on pre-existing radiation that passes through the region where the transient events happen. A (direct) problem of
electromagnetic field transformation was solved as a result of the flash transient
events and have provided a recipe for observers in Section 5.5 of Vorgul and Helling [2015]. 
The main findings are:

(a) The transformation of coherent emission by fast (compared to hydrodynamic processes in the atmosphere) transient
processes in the medium of its propagation can be observable.

(b) Attributing larger signal perturbations to more powerful processes
is not always correct, as more powerful processes can lead to smaller
response signals.

(c) Flash ionisation processes of intermediate intensity can be
detected by short-term amplification of the original, unperturbed
signal. The response field’s duration can be related to the flash
ionization’s time-scale and intensity.

(d) If the initial emission signal is powerful enough to
detect the character of damping,  small pulse’s amplitude and
duration in the response field can provide information about the
ionization flash magnitude (peak value of conductivity) and duration.

(e) If the detection of the electromagnetic field’s variation is
possible for several related flashes (like multiple strokes of the same
lightning discharge) and allows deriving the conductivity peak values
from additional information, it would be possible to determine whether
the discharges started as runaway electrons breakdown or rather like
capacitor discharges. This can be done by deriving the dependence of
the field responses’ maximum values on the maximum conductivity
change. The absence of a noticeable plateau in this dependence [see
Fig. 5 in Vorgul and Helling, 2015] suggests that the runaway breakdown
discharge is likely to be a source of the ionization, which requires
about 10 times lower voltage to enable the discharge in comparison to
a conventional capacitor--like discharge. If the plateau is noticeable,
then a capacitor--like discharge is more probable.

\section{Conclusion}

Extrasolar planets and brown dwarfs are exposed to a variety of ionising processes, externally, globally and locally. Depending on their location in the galaxy (star forming region vs. galactic bulge) but also on the kind of host star or companions, their atmospheres will be more or less dynamic or chemically altered by plasma processes. Both object classes (exoplanets and brown dwarfs) do form clouds  that can be very different, chemically and dynamically, from what we know from Earth or in the solar system. While this results in different cloud particles sizes and material compositions, the basic processes of lightning initiation and propagation are fundamentally the same. We therefore can apply concepts developed on Earth in order to study lightning occurring in extraterrestrial planets. Predicting spectral fingerprints will, however, require the modelling of the actual gas composition which can be very different from Earth.

\textbf{Acknowledgements.} Ch.~H. and I.~V. highlight the financial support of the European community under the FP7 ERC starting grant 257431. Most of the literature search has been performed using ADS.

\section*{References}
\everypar={\hangindent=1truecm \hangafter=1}
	
Ardaseva, A., Rimmer, P. B., Waldmann, I., Rocchetto, M., Yurchenko, S. N., Helling, Ch., Tennyson, J., Lightning chemistry on Earth-like exoplanets, \textit{MNRAS}, \textbf{470}, 187, 2017
	
Burinskaya, T. M. and Shevelev, M. M., Generation of auroral kilometric radiation in inhomogeneous magnetospheric plasma, \textit{Geomagnetism and Aeronomy}, \textbf{57}, 16, 2017 
	
Casewell, S.\,L., K.\,A.~Lawrie, P.\,F.\,L.~Maxted, M.\,S.~Marley, J.\,J.~Fortney, P.\,B.~Rimmer, S.\,P.~Littlefair, G. Wynn, M.\,R.~Burleigh, and Ch. Helling, Multiwaveband photometry of the irradiated brown dwarf WD0137-349B, \textit{MNRAS}, \textbf{447} (4), 3218, 2015.

Cairns, R. A.; Vorgul, I.; Bingham, R., Cyclotron Maser Radiation from an Inhomogeneous Plasma, \textit{Physical Review Letters}, \textbf{101}, 215003, 2008 

Cairns, R. A.; Vorgul, I.; Bingham, R.; Ronald, K.; Speirs, D. C.; McConville, S. L.; Gillespie, K. M.; Bryson, R.; Phelps, A. D. R.; Kellett, B. J. et al., Cyclotron maser radiation from inhomogeneous plasmas, \textit{Physics of Plasmas}, \textbf{18}, 022902, 2011

{{Gray}, M.},   Maser Sources in Astrophysics, Cambridge University Press, 2012

Haldoupis, C., Neubert, T., Inan, U. S., Mika, A., Allin, T. H., Marshall, R. A., Subionospheric early VLF signal perturbations observed in one-to-one association with sprites, \textit{JGRA}, \textbf{109}, A10303, 2004 

Hallinan, G., Antonova, A., Doyle, J. G., Bourke, S., Brisken, W. F., Golden, A., Rotational Modulation of the Radio Emission from the M9 Dwarf TVLM 513-46546: Broadband Coherent Emission at the Substellar Boundary?, \textit{ApJ}, \textbf{653}, 690, 2008 

Hallinan, G., S.\,P.~Littlefair, G.~Cotter, S.~Bourke, L.\,K.~Harding,
J.\,S.~Pineda, R.\,P.~Butler, A.~Golden, G.~Basri, J.\,G.~Doyle,
M.\,M.~Kao, S.\,V.~Berdyugina, A.~Kuznetsov, M.\,P.~Rupen, and
A. Antonova, Magnetospherically driven optical and radio aurorae at the end of the stellar main sequence,  \textit{Nature}, \textbf{523}, 568, 2015.

Helling, Ch., P.\,B.~Rimmer, I.\,M.~Rodriguez--Barrera, K. Wood, G.\,B.~Robertson, and C.\,R.~Stark, 
Ionisation and discharge in cloud--forming atmospheres of brown dwarfs and extrasolar planets, 
\textit{PPCF}, \textbf{58} (7), 4003, 2016a.

Helling, Ch., R.\,G.~Harrison, F. Honary, D.\,A.~Diver, K.~Aplin, I.~Dobbs--Dixon, U.~Ebert, S.~Inutsuka, F.\,J. Gordillo--Vazquez, and S.~Littlefair, Atmospheric electrification in dusty, reactive gases in the solar system and beyond, \textit{Surv. in Geophy.}, \textbf{37} (4), 705, 2016b.

Helling, Ch., P.~Woitke, and W.--F.~Thi, Dust in brown dwarfs and extra-solar planets. I. Chemical composition and spectral appearance of quasi-static cloud layers, \textit{Astron. Astrophys.}, \textbf{485}, 547, 2008.

Hodos\'an,~G., P.\,B.~Rimmer, and Ch.~Helling, Is lightning a possible source of the radio emission on HAT-P-11b? \textsl{MNRAS}, \textbf{461}, 1222--1226, 2016.

Inan, U. S., Cummer, S. A., Marshall, R. A., A survey of ELF and VLF research on lightning-ionosphere interactions and causative discharges, \textit{JGRA}, \textbf{115}, E36

Lee, G., I.~Dobbs--Dixon, Ch.~Helling, K.~Bognar, and P.~Woitke, Dynamic mineral clouds on HD 189733b. I. 3D RHD with kinetic, non-equilibrium cloud formation, \textit{Astron. Astrophys.}, \textbf{594}, 48, 2016.

Longstaff, E. S.,  Casewell, S. L., Wynn, G. A.,  Maxted, P. F. L.,  Helling, Ch., Emission lines in the atmosphere of the irradiated brown dwarf WD0137-349B, \textit{MNRAS}, \textbf{471}, 1728, 2017

Nikolov, N., D.\,K.~Sing, N.\,P.~Gibson, J.\,J.~Fortney, T.\,M.~Evans, J.\,K.~Barstow, T.~Kataria, and P.\,A.~Wilson, VLT FORS2 comparative transmission spectroscopy: Detection of Na in the atmosphere of WASP-39b from the ground, \textit{ApJ}, \textbf{832}, 191, 2016.

Schmidt, S.\,J., S.\,L.~Hawley, A.\,A.~West, J.\,J.~Bochanski, J.\,R.\,A.~Davenport, J.~Ge, and D.\,P.~Schneider, BOSS ultracool dwarfs. I. Colors and magnetic activity of M and L dwarfs, \textit{AJ}, \textbf{149}, 158, 2015.
	
Schellart, P., T.\,N.\,G.~Trinh, S.~Buitink, et al., Probing atmospheric electric fields in thunderstorms through radio emission from cosmic--ray--induced air showers, \textit{Phys. Rev. Lett.}, \textbf{114}, 165001, 2015.

Sing, D., J.\,J.~Fortney, N.~Nikolov, H.\,R.~Wakeford, T.~Kataria, T.\,M.~Evans, S.~Aigrain, G.\,E.~Ballester, A.\,S.~Burrows, D.~Deming, J.--M.~D\'esert, N.\,P.~Gibson, G.\,W.~Henry, C.\,M.~Huitson, H.\,A.~Knutson, A. Lecavelier Des Etangs, F.~Pont, A.\,P.~Showman, A.~Vidal--Madjar, M.\,H.~Williamson, and P.\,A.~Wilson, A continuum from clear to cloudy hot--Jupiter exoplanets without primordial water depletion, \textit{Nature}, \textbf{529}, 59, 2016.

Pritchett P. L., R. Strangeway, R. Ergun, and C. Carlson., Generation and propagation of cyclotron maser emissions in the finite auroral kilometric radiation source cavity, \textit{J. Geophys. Res.}, \textbf{107}, A12, 2002

Rimmer, P. B., Ch.~Helling, and C.~Bilger, The influence of galactic cosmic rays on ion--neutral hydrocarbon chemistry in the upper atmospheres of free-floating exoplanets, \textit{IJAsB}, \textbf{13}, 173, 2014.

Rodriguez--Barrera, M.\,I., Ch.~Helling, C.\,R.~Stark, and A.\,M.~Rice, Reference study to characterize plasma and magnetic properties of ultracool atmospheres, \textit{MNRAS}, \textbf{454}, 3977, 2015.

Rodriguez--Barrera, M.\,I., Ionisation in atmospheres of  brown dwarfs ad giant gas planets, \textit{PhD Thesis}, University of St Andrews, 2016.

Williams, P. K. G., S.\,L.~Casewell, C.\,R.~Stark, S.\,P.~Littlefair, Ch.~Helling, and E.~Berger, The first millimeter detection of a non-accreting ultracool dwarf, \textit{MNRAS}, \textbf{815}, 64, 2015.
	
Vorgul, I., and Ch. Helling, Flash ionization signature in coherent cyclotron emission from brown dwarfs, \textit{MNRAS}, \textbf{ 458}, 1041, 2016.

\end{document}